\documentclass[submission,copyright,creativecommons]{eptcs}
\providecommand{\event}{MARS 2017} 

\usepackage{tikz}
\usepackage{amssymb}
\usepackage{subcaption}
\usepackage{mathtools}
\usepackage{graphicx}
\usepackage{array}
\usetikzlibrary{automata,positioning}
\usepgflibrary{arrows}
\usetikzlibrary{graphs}
\usepackage{float}
\usepackage{url}

\newenvironment{mcrl2}%
{\begin{trivlist}\item\begin{tabular}{@{}>{\bf}p{2em}L@{\ }L@{\ }L@{\ }L@{\ }L@{\ }L@{\ }L@{\ }L}}%
{\end{tabular}\end{trivlist}}

\newcolumntype{L}{>{$}l<{$}}
\newcolumntype{C}{>{$}c<{$}}
\newcolumntype{R}{>{$}r<{$}}

\title{Modelling and Verification of\\ a Cluster-tree Formation Protocol Implementation\\ for the 
IEEE 802.15.4 TSCH MAC Operation Mode}
\author{
Mahmoud Talebi \qquad\quad Jan Friso Groote
\institute{\footnotesize\sl Departement of Mathematics and Computer Science\\
Eindhoven University of Technology, The Netherlands}
\email{\footnotesize \sl Email: \tt \{M.Talebi, J.F.Groote\}@tue.nl}
\and
Conrad Dandelski
\institute{\footnotesize\sl Nimbus Centre for Embedded Systems Research\\ 
Cork Institute of Technology, Ireland}
\email{\footnotesize \sl Email: \tt conrad.dandelski@mycit.ie}
}

\date{}

\def\titlerunning{Modelling and Verification of a Cluster-tree Formation Protocol for IEEE 802.15.4 TSCH}
\def\authorrunning{M. Talebi, J.F. Groote \& C. Dandelski}

\begin{document}
\maketitle

\begin{abstract}
\vspace{-0.2cm}
\noindent Correct and efficient initialization of wireless sensor networks can be challenging in the face of many uncertainties 
present in ad hoc wireless networks. In this paper we examine an implementation for 
the formation of a cluster-tree topology in a network
which operates on top of the TSCH MAC operation mode of the IEEE 802.15.4 standard, and investigate it using formal methods. 
We show how both the mCRL2 language and toolset help us in identifying 
scenarios where the implementation does not form a proper topology. More importantly, our analysis
leads to the conclusion that the cluster-tree formation algorithm has a super linear time complexity. So, it does not
scale to large networks. 
\end{abstract}
\vspace{-0.4cm}
\section{Introduction}
The IEEE 802.15.4 protocol standard is used for low-rate and low-powered Personal Area Networks (PANs). 
The IEEE 802.15.4e standard was released in 2012 to extend the IEEE 802.15.4 protocol to overcome limitations in performance and reliability~\cite{TSCH2012,802.15.4e-survey}. This extension was subsequently incorporated into the IEEE 802.15.4 standard in 2015, and among other changes introduces the Time-Slotted Channel Hopping (TSCH) MAC operation mode, with the goal of ensuring reliability by providing more structure to wireless communication, as well as scalability and robustness by channel hopping~\cite{TSCH2015}. 

In the TSCH mode, robustness against noise and interference is provided by a mechanism for channel hopping in consecutive 
units of time called time-slots, and the reliability is provided by communicating according to a deterministic schedule over 
time-slots. The schedule is created by higher-layer protocols on the PAN coordinator, which is also in charge of initiating and maintaining the PAN.

For multi-hop applications, a logical structure is maintained which is called a topology, and is often a cluster-tree. This structure is used for multi-hop routing and often also for creating  schedules. However, as networks become larger and/or denser the problem of designing an algorithm for forming the cluster-tree efficiently becomes non-trivial, and ensuring its correctness becomes difficult.

A problem is that the IEEE 802.15.4 protocol standard is not explicit about many of the operational details of the protocol, 
including the cluster-tree formation algorithm. Therefore, we used the description of an early implementation~\cite{dandelski2016rll}
and investigated its correctness using the mCRL2 toolset \cite{mCRL2,groote2014modeling}. 
Our findings were in line with results of simulations by the designers, with substantial saving of time in case of verification 
compared to simulations. Some, but not all, observations were incorporated in later implementations.

In order to investigate the implementation, 
we first model the lower layers (including the IEEE 802.15.4 TSCH mode protocol) in mCRL2. 
This model is presented in Section~\ref{sec:twolayer}. 
We then model the cluster-tree formation protocol operating on TSCH. This model is quite large and can 
be found in the model repository. We formally analyse correctness using the 
mCRL2 model checker. We subsequently present a full argument on the scalability of the algorithm, based on the results 
derived by verification. 

From our analysis we derived a number of guidelines on how to design a better cluster-tree formation protocol (in Section~\ref{sec:conc}). 
But we do not devise such an alternative protocol, as the purpose of this paper is to show that with relatively little effort
formal methods can be of great help in getting insight in the behaviour of protocols used in the realm of wireless networks. 

\section{Related work}
When discussing cluster-trees, we are essentially referring to spanning trees, in which each node is a vertex in the tree, 
and for every pair of node identifiers ($A$,$B$) there is a unique path between the node with identifier $A$ and the node with 
identifier $B$ in the tree. The problem of forming a cluster-tree is directly related to the distributed minimum 
spanning tree (MST) algorithms in an asynchronous setting~\cite{lynch1996distributed}. However, several assumptions by distributed 
MST algorithms are in conflict with the situation in wireless sensor networks: links between nodes should be reliable meaning that 
no collisions or message losses occur, and processes on nodes should have complete knowledge regarding the neighbours of the 
node. Moreover, 
in some versions of the distributed spanning tree algorithm the leader or the root node is elected during the course of 
execution~\cite{gallager1983distributed,garay1998sublinear}, while in the algorithms we consider the root node is fixed.

The general strategy in distributed spanning tree algorithms is to start with a spanning forest, and consistently merge trees until eventually a single spanning tree is formed~\cite{lynch1996distributed}. However, in IEEE 802.15.4 due to having a single 
initiator an essentially different approach is taken which forms a tree by a series of waves (i.e., floodings) sent through 
the network. Therefore, the algorithm presented here fits the definition of echo algorithms in the theory of distributed algorithms~\cite{fokkink2013distributed}.

To our knowledge there have been relatively few attempts to rigorously verify the IEEE~802.15.4 protocol up until now. Recently, in~\cite{kauer2016formal} the authors used the UPPAAL framework to analyse inconsistencies in slot allocation in an implementation of the distributed synchronous multi-channel extension (DSME) mode of the IEEE 802.15.4 MAC layer protocol. Earlier, in~\cite{gross2007does} the authors modelled a part of the IEEE 802.15.4 specification using the MODEST language to investigate the influence of clock precision on energy consumption. In any case, this work is one of the first to formally analyse parts of the TSCH operation mode.

\section{Modelling the TSCH MAC operation mode}\label{sec:tsch}
In this section we introduce the service of the TSCH MAC layer as specified in~\cite{TSCH2015}. 
We also sketch how we model protocols using this MAC protocol. We leave out the details of the channel hopping procedure from 
the TSCH MAC layer. 
The channel hopping procedure is an essential part of TSCH, which provides scalability and reliability in communication. However, as the correctness of channel hopping procedures is not a 
concern in this paper we abstract away all related concepts.

In what follows we first give an informal specification for the TSCH MAC operation mode in Section~\ref{sec:tsch}. Then, in Section~\ref{sec:twolayer} we show how the informal descriptions can be encoded formally.

\subsection{The TSCH MAC operation mode}\label{sec:tsch}
In IEEE 802.15.4, 16 channels are available for communication. The TSCH MAC operation mode is time-slotted, and divides time 
into equally sized chunks called {\it time-slots}. For each time-slot on each channel a node can be either a receiver, 
a transmitter, or not involved (e.g., if it is sleeping). 

In TSCH, time-slots have two types: they are either {\it dedicated}, meaning that only one transmitter and one or more receivers 
are involved, or they are {\it shared}, meaning that there can be zero, one or more transmitters attempting to transmit a frame and multiple receivers listening. In case of shared time-slots conflicts between transmitters are resolved using CSMA-CA.

For communicating over time-slots, nodes must share a common schedule, which is designed such that their frames do not collide. Otherwise, collisions are a possibility. Let us assume that in a time-slot node $A$ is the transmitter (TX) and node $B$ is the 
receiver (RX), according to the schedule. The events occurring in the time-slot are as follows: at the beginning of the time-slot 
both $A$ and $B$
wait for a time period. After waiting, first $B$ starts listening on the channel, then $A$ becomes active and transmits its 
message. When $A$ is done transmitting the message, it stops and waits for some time. Next, $B$ stops listening and waits. After 
some time node $A$ starts listening on the channel. Then $B$ becomes active and if it has succeeded in receiving the frame, it transmits an acknowledgement. If not, it does no action. After a while (before the start of the next time-slot) $A$ stops 
listening. In this way by the end of the time-slot,
$A$ and $B$ reliably know the outcome of the communication. Also note that listening 
starts before the expected transmission time and ends sometime after to allow possible synchronisation imperfections.

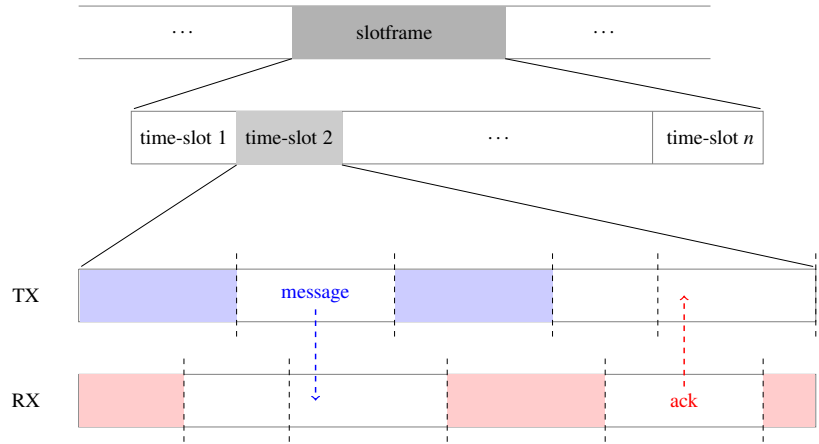
\begin{figure}
\centering
\scalebox{0.7}{
\begin{tikzpicture}
\def\x0{-1}

\draw[gray,very thin] (0,-1) -- (12,-1);
\draw[gray,very thin] (0,-2) -- (12,-2);

\fill[black!30!white] (4.05,-1) rectangle (8.1,-2);

\node (a) at (6,-1.5){slotframe};
\node (a) at (2,-1.5){\ldots};
\node (a) at (10,-1.5){\ldots};

\draw (4.05,-2) -- (1.05,-2.95) node{};
\draw (8.1,-2) -- (12.95,-2.95) node{};

\draw[gray,very thin] (1,-4) -- (13,-4) -- (13,-3) -- (1,-3) -- (1,-4);
\draw[gray,very thin] (3,-3) -- (3,-4);
\draw[gray,very thin] (5,-3) -- (5,-4);
\draw[gray,very thin] (10.9,-3) -- (10.9,-4);

\fill[black!20!white] (3,-3) rectangle (5,-4);

\node (a) at (2,-3.5){time-slot~1};
\node (b) at (4,-3.5){time-slot~2};
\node (x) at (8,-3.5){\ldots};
\node (x) at (12,-3.5){time-slot~$n$};

\draw (3,-4.02) -- (0.05,-5.95) node{};
\draw (5,-4.02) -- (13.95,-5.95) node{};

\node (c) at (-1,-6.5){TX};
\draw[gray,very thin] (0,-6) -- (14,-6) -- (14,-7) -- (0,-7) -- (0,-6);

\fill[blue!20!white] (0.02,-6.02) rectangle (3,-7);
\draw[thin,dashed] (3,-5.7) -- (3,-7.3);

\fill[blue!20!white] (6,-6.02) rectangle (9,-7);
\draw[thin,dashed] (6,-5.7) -- (6,-7.3);
\draw[thin,dashed] (9,-5.7) -- (9,-7.3);
\draw[thin,dashed] (11,-5.7) -- (11,-7.3);

\draw[thin,dashed] (14,-5.7) -- (14,-7.3);

\node[blue] (l) at (4.5,-6.5){message};

\draw[dashed,->,blue,thick] (l) -- (4.5,-8.5) node{};

\node (d) at (-1,-8.5){RX};
\draw[gray,very thin] (0,-8) -- (14,-8) -- (14,-9) -- (0,-9) -- (0,-8);

\fill[red!20!white] (0.02,-8) rectangle (2,-9);
\draw[thin,dashed] (2,-7.7) -- (2,-9.3);
\draw[thin,dashed] (4,-7.7) -- (4,-9.3);

\fill[red!20!white] (7,-8) rectangle (10,-9);
\draw[thin,dashed] (7,-7.7) -- (7,-9.3);
\draw[thin,dashed] (10,-7.7) -- (10,-9.3);

\fill[red!20!white] (13,-8) rectangle (13.98,-9);
\draw[thin,dashed] (13,-7.7) -- (13,-9.3);

\node[red] (j) at (11.5,-8.5){ack};

\draw[dashed,->,red,thick] (j) -- (11.5,-6.5) node{};

\end{tikzpicture}}
\caption{A slotframe with $n$ time-slots and communication between a transmitter and a receiver.}
\label{fig:tsch}
\end{figure}

In TSCH, nodes have a common understanding of time, and agree on schedules. The schedules are repeated over 
slotframes, which are a collection of time-slots. A summary of these concepts and their relations is given in Figure~\ref{fig:tsch}.

\subsection{The two layered model}\label{sec:twolayer}

In this section we outline the formal specification of the general behaviour of the network as a two layered model, with one 
representing the physical and the TSCH operation mode of the MAC layer of the protocol stack, and the other representing the 
upper layers. At the same time, we also introduce the mCRL2 formal specification language for specifying and verifying the 
behaviour of concurrent systems~\cite{mCRL2}. 

For the specification of the behaviour the following data types are needed. Let {\it Message\_Type} be a sort which at least contains the element \texttt{EMPTY}. The inclusion of the element \texttt{EMPTY} is due to the fact that in our approach in modelling broadcasting, in each time-slot we require every node to submit a message to the physical layer. Define:

\begin{mcrl2}     
sort & {\it Message} & = {\bf struct}~{\it message}({\it src}:\mathbb{N},{\it channel}:\mathbb{N},{\it type}:{\it Message\_Type},{\it data}:{\it List}(\mathbb{N}));\\
 & {\it Traffic} &= \mathit{List}({\it Message});
\end{mcrl2}

The above ${\bf sort}$ statement defines custom data types. The sort ${\it Message}$ stands for a structured sort, which can be constructed using the function ${\it message}$ applied to the elements that follow. The element ${\it src}$ in ${\it message}$ is the identifier of the transmitter of the message which belongs to the set of natural numbers, {\it channel} is the channel over which the message is transmitted, again belonging to the set of natural numbers, {\it type} is the type of the message, and \textit{data} is a list of data carried by the message (payload), which belongs to the set of lists (sequences) of natural numbers.

The sort $\mathit{Traffic}$ is a list of messages. 
This models all the messages coming from all the sources travelling over all the channels at a certain point in time in the network. 

Constants and functions are defined using the ${\bf map}$ statement:
\begin{mcrl2}     
map & \texttt{MAX\_ID} : \mathbb{N};\\
& \texttt{NUM\_OF\_CHANNELS} : \mathbb{N};\\
\end{mcrl2}
The above statement defines two constants with values from the set of natural numbers, as mappings with zero arguments. We use the constant $\texttt{MAX\_ID}$ to indicate that $1,\ldots,\texttt{MAX\_ID}$ are identifiers of processes in the network, and 
the constant $\texttt{NUM\_OF\_CHANNELS}$ indicates that the channels available are $1,\ldots,\texttt{NUM\_OF\_CHANNELS}$. 

Based on the informal specification given in Section~\ref{sec:tsch}, the behaviour of the network depends on the type of time-slots (dedicated or shared). Here, we only define the behaviour for dedicated time-slots.

We model the physical and MAC layers of the network in a general fashion, such that it applies to any protocol implemented over the dedicated time-slots of TSCH. This is done using the process {\it TSCH}. The main purpose of defining this process is to model the peculiarities of phenomena such as {\it broadcasting} and {\it collisions} using mCRL2. 

We model the transmission and receiving of messages using the following ${\bf act}$ statement:
\begin{mcrl2}
act&{\it corecv},{\it recv},{\it received}&:{\it Message};\\
&{\it cosend},{\it send},{\it sent}&:{\it Message};\\
\end{mcrl2}

Actions represent events occurring in the system, and are the building blocks in defining the behaviour of processes. 
The action $\mathit{received}$ indicates that a message has been received in the MAC layer and is handed over to the upper layer.
Concretely, the MAC layer performs a $\mathit{corecv}$ action and the upper layer performs a $\mathit{recv}$ which synchronise into
the $\mathit{received}$ action. Similarly, the action $\mathit{sent}$ means that the upper layer indicates that the MAC layer must
send a message. This action is the result of the synchronisation of actions $\mathit{send}$ by the upper layer and $\mathit{cosend}$ by
the MAC layer. 

The synchronisations are made explicit by a communication set $C$:
\[C=\{{\it recv}|{\it corecv}\rightarrow {\it received},~{\it send}|{\it cosend}\rightarrow {\it sent}\}.\]
A synchronisation can only take place when both constituent actions are ready and have the same data of type message. 
The communication operator $\Gamma_C$ is used in the process to indicate that these communications are possible.
The operator $\nabla_V$ only allows actions in the set $V$ to happen. It can be used to enforce communications
by only allowing the result of a synchronisation and not the constituent actions. 
Concretely, when applying $\nabla_{\{\mathit{sent},\mathit{received}\}}$ the result is that actions 
${\it cosend}$ and ${\it send}$ cannot be performed independently and must communicate to ${\it sent}$. The same applies
to $\mathit{corecv}$, $\mathit{recv}$ and $\mathit{receive}$. 

Assuming that the process ${\it TSCH}$ describes the physical and MAC layers and the processes ${\it Node}$ describe the upper layer, 
a TSCH network with $n=\texttt{MAX\_ID}$ nodes is specified by the process ${\it Network}$:

\begin{mcrl2}
proc&\mathit{Network}=\nabla_{\{{\it sent},{\it received}\}}\Gamma_C({\it TSCH}\parallel {\it Node}(1,c_1,d_1)\parallel\ldots\parallel {\it Node}(n,c_n,d_n));&\\
\end{mcrl2}

In the following two sections, we give the details on how processes ${\it TSCH}$ and ${\it Node}$ are specified.

\subsubsection{Specification of the {\it TSCH} process}\label{sec:physmac}
In this section, we specify the behaviour of the physical and the MAC layers using the mCRL2 language, as the process ${\it TSCH}$.
We start out with the $\mathit{TSCH}$ process.

We first define the following functions which enable us to manipulate the traffic in the network:
\begin{mcrl2}     
map & {\it findId} & : \mathbb{N}\#{\it Traffic}\rightarrow {\it Message};\\
& {\it findChannel} & : \mathbb{N}\#{\it Traffic}\rightarrow {\it Traffic};
\end{mcrl2}
The function $\mathit{findId}$ accepts a node identifier and a list of messages ($\mathit{Traffic}$), and returns the first 
message from a node with the mentioned identifier in this list, only if such a message exists. The function $\mathit{findChannel}$ 
accepts a channel number and a list of messages, and returns a list of messages sent on a certain channel. We specify the 
behaviour of ${\it TSCH}$ 
using two processes ${\it TSCH}_1$ and ${\it TSCH}_2$. 

The process ${\it TSCH}_1$ represents the first stage of the events on the physical layer. 
In this stage, $\mathit{TSCH}_1$ collects from each node $i$ a message to be sent (using the action $\mathit{cosend}$) 
and adds it to the sent transmissions $\mathit{str}$ such that it will be transmitted in the second stage.

\begin{mcrl2}
proc&{\it TSCH}_1(i: \mathbb{N},str:{\it Traffic})=&\\
&\quad (i\leq \texttt{MAX\_ID})&\\
&\quad\quad\quad\quad \rightarrow~\sum_{c: \mathbb{N},d:{\it List(\mathbb{N})}}{\it cosend}({\it message}(i,c,d))
\cdot {\it TSCH}_1(i+1,str\lhd{\it message}(i,c,d))&\\
&\quad\quad\quad\quad \,\Diamond\,\,\,\,~ {\it TSCH}_2(1,str,{\it removeCollision}(1,str));&\\
\end{mcrl2}
The notation $c{\rightarrow} p \Diamond q$ stands for the if-then-else. The operator $\lhd$ is used to put an element at the end of a list.
Note that every node should send a message. Nodes which are only listening, communicate 
\texttt{EMPTY} messages to the TSCH in this stage. 

While transmitting messages, collisions may occur on the physical layer. A collision 
occurs over a channel $c$ if there are at least two messages $m_1$ and $m_2$ which are on the same channel $c$. 
In case of a collision on channel $c$, none of the messages sent over channel $c$ are readable and therefore they
need to removed when the sent transmissions become the received transmissions in $\mathit{TSCH}$. This behaviour is
described using a function ${\it removeCollision}$:
\begin{mcrl2}
map & {\it removeCollision} : \mathbb{N}\#{\it Traffic}\rightarrow {\it Traffic};\\
var & \mathit{tr}_1 : {\it Traffic}; ~~\mathit{chann} : \mathbb{N};\\
eqn & {\it removeCollision}(\mathit{chann},\mathit{tr}_1)=\\
&\quad\quad\quad\quad{\it if}(\mathit{chann}\leq \texttt{NUM\_OF\_CHANNELS},\\
&\quad\quad\quad\quad\quad\quad\quad {\it if}(\#(findChannel(\mathit{chann},\mathit{tr}_1))\approx 1,\\
&\quad\quad\quad\quad\quad\quad\quad\quad\quad\quad\quad {\it removeCollision}(\mathit{chann}+1,\mathit{tr}_1)\lhd {\it head}({\it findChannel}(\mathit{chann},\mathit{tr}_1)),\\
&\quad\quad\quad\quad\quad\quad\quad\quad\quad\quad\quad {\it removeCollision}(chann+1,\mathit{tr}_1)\lhd {\it message}(0,chann,\texttt{EMPTY},0)),\\
&\quad\quad\quad\quad\quad\quad\quad[]);
\end{mcrl2}

The function ${\it removeCollision}$ processes all the messages in the traffic over all channels, and for each channel determines which message is broadcast. The message is then added to the list of results. In case of collision on a channel $c$, an \texttt{EMPTY} message over $c$ is added to the resulting list.

In the second stage, the process ${\it TSCH}_2$ checks the channel each node $i$ is operating on using the expression 
$channel({\it findId}(i,str))$, and picks a message from that channel in the list $rtr$ and delivers it to the node.

\begin{mcrl2}
proc&{\it TSCH}_2(i: \mathbb{N},str:{\it Traffic},{\it rtr}:{\it Traffic})=&\\
&\quad\quad\quad {\it corecv}({\it head}({\it findChannel}(channel({\it findId}(i,str)),rtr)))\cdot\\
&\quad\quad\quad\quad\quad\quad ((i< \texttt{MAX\_ID})\rightarrow
{\it TSCH}_2(i+1,str,rtr)
 \,\,\Diamond\,\,
 {\it TSCH}_1(1,[])
);\\
\end{mcrl2}
Upon finishing this stage, the behaviour changes back to process ${\it TSCH}_1$. The process ${\it TSCH}$ which represents the complete behaviour of the lower layers is then defined as:
\begin{mcrl2}
proc&{\it TSCH}={\it TSCH}_1(1,[]);&\\
\end{mcrl2}

\subsubsection{General specification of process {\it Node}}\label{sec:node}

The general structure of the processes ${\it Node}$ is as follows:

\begin{mcrl2}
proc&{\it Node}({\it id}:\mathbb{N},{\it ch}:\mathbb{N},{\it d}:{\it Data})=&\\
&\quad\quad\sum_{i\in I} \gamma_i(d)\rightarrow \Big({\it send}({\it message}(id,ch,t_i,g_i(id,d))\cdot &\\
&\quad\quad\quad\quad\quad\quad\quad\quad\sum_{id': \mathbb{N},t':\mathit{Message\_Type},d':{\it Data}}(id'\leq\texttt{MAX\_ID})\rightarrow &\\
&\quad\quad\quad\quad\quad\quad\quad\quad\quad{\it recv}({\it message}(id',{\it ch},t',d'))\cdot {\it Node}({\it id},\sigma_i({\it ch},d),\delta_i({\it ch},d))\Big)&\\
&\quad\quad+&\\
&\quad\quad\sum_{j\in J} \gamma_j(d)\rightarrow \Big({\it send}({\it message}(id,{\it ch},\texttt{EMPTY},[])\cdot &\\
&\quad\quad\quad\quad\quad\quad\quad\quad\sum_{id': \mathbb{N},t':\mathit{Message\_Type},d':{\it Data}}(id'\leq\texttt{MAX\_ID})\rightarrow &\\
&\quad\quad\quad\quad\quad\quad\quad\quad\quad{\it recv}({\it message}(id',{\it ch},t',d'))\cdot {\it Node}({\it id},\sigma'_j({\it ch},d,id',d'),\delta'_j({\it ch},d,id',d'))\Big)&\\
\end{mcrl2}
\noindent%
In the process $\mathit{Node}$, the number $\mathit{id}$ is the identifier of the node. 
The parameter $\mathit{ch}$ is the channel the node operates on. The parameter $\mathit{d}$ of sort $\mathit{Data}$
is a placeholder for the rest of the state of a node. In the concrete description of $\mathit{Node}$ $d$ is replaced
by $7$ concrete parameters. 

Each process $\mathit{Node}$ represents a 
wireless node with a single antenna which at any time can only be in a transmit mode or a receive mode, based on which the 
behaviour of a node is split into two main parts, represented by the two summations. 

The set $I$ is a finite set of indices $i$ of all conditions $\gamma_i$ under which the custom-defined protocol transmits a message. 
Similarly, the set $J$ is the set of indices $j$ of all conditions $\gamma_j$ under which the custom-defined protocol waits for a message. 
The first summation 
(choosing over set $I$) represents the behaviour of nodes in the transmit mode. In this mode, if condition $\gamma_i$ with 
data $d$ holds, the node sends a message of type $t_i$ over ${\it ch}$, containing data $g_i(id,d)$. The antenna 
is passively exposed to any signal occurring over channel ${\it ch}$ and this is reflected by the subsequent {\it recv} action, which 
records the messages broadcast at the same time over the operating channel. However, being in transmit mode, the node cannot 
react to this message. Therefore the values of functions $\sigma_i$ and $\delta_i$ which determine the next state of the node 
are not affected by the received data.

The second summation (choosing over set $J$) shows the behaviour of nodes in the receive mode. In this mode, a node does not transmit any messages. This particular behaviour is shown by a ${\it send}$ action of a message of type \texttt{EMPTY} over channel $\mathit{ch}$. The ${\it send}$ action is only to inform the ${\it TSCH}$ process of the value of ${\it ch}$. The main part of the behaviour is then expressed by the ${\it recv}$ action. Functions $\sigma'_i$ and $\delta'_i$ express the change to the state of the node based on the received data and the previous state.

Note that despite the fact that two communications occur in each time-slot in TSCH, namely transmission of a frame and its acknowledgement, we do not explicitly enforce them in our model. Therefore the user defined protocol may choose to  include the acknowledgement in its behaviour by specifying two rounds of communication per time-slot, or to leave the acknowledgements out.

\section{The cluster-tree formation protocol}\label{sec:ver1}
In this section, we specify the cluster-tree formation protocol using mCRL2, and then we present the results of the verification of this protocol which shows the flaws in its design.
\subsection{Informal specification}\label{sec:informal}
A slotframe in the cluster-tree formation protocol consists of 12 time-slots, out of which two are reserved for the operation of the cluster-tree formation protocol. The idea is that the protocol establishes and maintains a cluster-tree topology in these two time-slots while the network is doing its normal operation in the rest of the time-slots.

Nodes can have one of 4 types. They can be either a cluster head (CH), a cluster slave (CS), a tentative cluster head (TCH), or a free node (FN). In the scenarios that we are interested in, the network starts with one cluster head node and the rest are free nodes. The original cluster head is a gateway node receiving commands from an external source, and it initiates the cluster-tree formation protocol.

The cluster-tree formation protocol takes a distance based approach. For each node, nodes in its broadcast range fall into two categories: close nodes, and other nodes in range. A node estimates the distance of another node based on its received signal strength (RSSI). Upon receipt of a message, a node measures the power of the signal, and compares it to a power threshold and determines the category of the transmitter.

Cluster heads in the network are connected to each other (they form a tree) in a setting in which every cluster head has one parent and a number of children. The original cluster head is an exception and has no parent. Based on this a cluster head is assigned a {\it tier}, which shows its distance from the original cluster head, with the original cluster head being assigned tier 0. In addition, each cluster head operates on only two channels: one to communicate with its parent cluster head, and one to communicate with its children. We call these channels the {\it parent channel} and the {\it assigned channel} respectively.

A cluster head broadcasts a \texttt{BEACON} message on its assigned channel to advertise its presence in the network, according to the scheme that follows. Cluster heads with even-numbered tiers advertise on the first time-slot reserved for cluster-tree formation in a slotframe, and cluster heads with odd-numbered tiers advertise on the second time-slot reserved for cluster-tree formation. Therefore every cluster head gets the chance to advertise once in each slotframe. The \texttt{BEACON} message includes the tier of the transmitter node.

The message is received by all the nodes which are within transmission range, and are listening on the assigned channel of the cluster head. When a free node receives the message, it estimates its distance from the transmitter, and if it is close then it sends an \texttt{ASSOCIATE} message in response and requests to become a cluster slave. A cluster slave operates on the same channel as its parent. Otherwise if the node is not close to the cluster head, it responds with a \texttt{BEACON\_ACK} to the request, and the cluster head then sends an \texttt{ACK\_RESPONSE} message which contains a \texttt{wait\_time}.

When an \texttt{ACK\_RESPONSE} message is received by a free node, it becomes a tentative cluster head. A tentative cluster head scans other channels one by one (staying in each for 2 time-slots) until either \texttt{wait\_time} expires, or it receives a \texttt{BEACON} from a cluster head which is close and has the same tier. In such a case the node becomes a cluster slave of the close cluster head, as previously explained. This is to achieve approximately the smallest tree in the network which covers all of the nodes.

If a tentative cluster head fails to find a close cluster head and \texttt{wait\_time} expires, it moves back to the channel of its parent and listens for a \texttt{BEACON} message. This time it sends an \texttt{ASSOCIATE} message in response and requests to become a child cluster head. The parent then forwards this message to the original cluster head. The original cluster head finds a suitable and possibly unoccupied channel, and sends an \texttt{ASSOCIATE\_ACK} message containing the assigned channel to the sender of the request, which is then forwarded to the tentative node, officially granting it the status of a cluster head.

We assume that the network is well connected, meaning that for each two nodes $A$ and $B$ in the network 
it is possible to deliver a message from $A$ to be $B$ in one or more hops. The main requirements that the above 
protocol should satisfy in such a network is the following:
\begin{itemize}
\item  Eventually every node in the network becomes either a cluster head or a cluster slave.
\end{itemize}

Below we present our method to check this property.

\subsection{Formal specification of the cluster-tree formation protocol in mCRL2}\label{sec:formalspec}
In this section we explain how the cluster-tree formation protocol introduced in Section \ref{sec:informal} can be formally specified. Based on the informal specification, we define data types {\it Node\_Type} and {\it Message\_Type} as follows:

\begin{mcrl2}     
sort & {\it Node\_Type}  = {\bf struct}~\texttt{CLUSTER\_HEAD}~|~\texttt{TENTATIVE}~|~\texttt{CLUSTER\_SLAVE}~|~\texttt{FREE};\\
 & {\it Message\_Type}  = {\bf struct}~\texttt{EMPTY}~|~\texttt{BEACON}~|~\texttt{BEACON\_ACK}~|~\texttt{ACK\_RESPONSE}~|\\
 &\hspace{9cm}~\texttt{ASSOCIATE}~|~\texttt{ASSOCIATE\_ACK};
\end{mcrl2}

Next, we model the physical topology of the network using the distance of nodes with respect to each other and by declaring the functions {\it inRange} and {\it inClose}.
\begin{mcrl2}     
map & {\it inRange} & : \mathbb{N}\#\mathbb{N}\rightarrow\mathbb{B};\\
& {\it inClose} & : \mathbb{N}\#\mathbb{N}\rightarrow\mathbb{B};\\
\end{mcrl2}
For two node identifiers $A$ and $B$ the value of the term $\mathit{inClose}(A,B)$ is true if and only if node $A$ is within range of node $B$, and is closer than the threshold. The term $\mathit{inRange}(A,B)$ is true, if and only if node $A$ is within range of node $B$ but farther away than the threshold.

Using these data types, and the general definition of a TSCH node, the basic declaration of the process modelling a node  is:
\begin{mcrl2}
proc&\mathit{Node}({\it id}:\mathbb{N},\mathit{ch}:\mathbb{N},\\
&\hspace*{1.8cm}type:{\it Node\_Type},{\it pid}:\mathbb{N},{\it pc}:\mathbb{N},{\it children}:{\it List}({\it Pair}),{\it tier}:\mathbb{N},state:\mathbb{N},{\it timeslot}:\mathbb{N})&\\
\end{mcrl2}
\noindent%
In this process, if a node is of type cluster head or cluster slave, $\mathit{pid}$ is the identifier of the parent of the node and $\mathit{pc}$ is the channel over which the parent of the node operates. Given type ${\it Pair}$ which defines any pair of natural numbers, a cluster head keeps a record of all its children and their assigned channels as a list of pairs of natural numbers. The parameter ${\it tier}$ is the tier of the cluster head or cluster slave and ${\it state}$ is the internal state of the node which in general holds information such as the node's history and what the node should do next. Finally, ${\it timeslot}$ is the information regarding the clock of the node, which is used by the node to keep its actions in sync with the rest of the network.

We model a TSCH network in which the original cluster head is initialised to the state:
\[\mathit{Node}(1,1,\texttt{CLUSTER\_HEAD},0,0,[],0,0,0),\]
and every other node with identifier $n\in\{2,\ldots,\texttt{MAX\_ID}\}$ is initialised to:
$\mathit{Node}(n,c_n,\texttt{FREE},0,0,[],0,0,0)$,
where $c_n$ is a channel number, chosen based on a scenario of interest. The process $\mathit{Network}$ given in Section \ref{sec:twolayer} then models a network of $n$ nodes running the cluster-tree formation protocol.

\subsection{Verification of the cluster-tree formation protocol in mCRL2}\label{section:verif}
We want to verify the property: ``eventually every node in the network is either a cluster head or a cluster slave''. In order to check the state of the whole network throughout its operation, we introduce a new action ${\it increase}$. In the specification of the process ${\it Node}$, every time the type of the node changes to \texttt{CLUSTER\_HEAD} or \texttt{CLUSTER\_SLAVE}, the action ${\it increase}$ is performed. Then the property ``eventually every node in the network is either a cluster head or a cluster slave'' can be expressed as ``for every possible computation at some finite point in the future the action ${\it increase}$ has occurred $\texttt{MAX\_ID}-1$ times''. This statement 
formulated in the modal $\mu$-calculus looks as follows:
\begin{equation}\label{eq:success}
\mu X(n:\mathbb{N}:=1).\left([{\it increase}] X(n+1)\wedge [\overline{{\it increase}}] X(n)\wedge \langle{\it true}\rangle{\it true}\right)\vee(n\approx \texttt{MAX\_ID})
\end{equation}

For a network of 3 nodes, we manually generated all the topologies in which the network is well connected. 
For all those cases, we verified the ${Network}$ process against property (\ref{eq:success}) using the mCRL2 model checker. 
The toolset generated a number of {\it counterexamples} for the property.

Below we present three 
basic scenarios in which the network does not behave as desired. In our 
representation of topologies that follow, we use dashed lines to show two nodes being in range 
and full lines to show two nodes being close. Moreover, we give the type of the nodes by annotating 
them by CH, CS, F and T which stand for cluster head, cluster slave, free and tentative respectively.

\vspace{2mm}
\noindent
\begin{minipage}[H]{0.72\textwidth}
{\bf Collision of acknowledgements}. In the cluster-tree formation protocol whenever a free node receives a \texttt{BEACON} message, it responds with an acknowledgement right away. However, in the physical topology given to the right, since the \texttt{BEACON} is sent by broadcast there could be more than one node responding on the same channel, and this leads to collision.
\end{minipage}
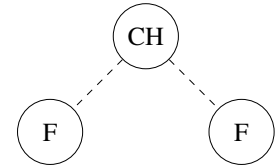
\begin{minipage}[H]{0.3\textwidth}
\centering
\scalebox{0.9}{
\begin{tikzpicture}[shorten >=1pt,node distance=2cm,on grid,auto]
   \node[state] (q_0)   {CH};
   \node[state] (q_1) [below right=of q_0] {F};
   \node[state] (q_2) [below left=of q_0] {F};
    \path[dashed,line width=0.4pt]
    (q_0) edge node {} (q_1)
          edge node {} (q_2);
\end{tikzpicture}
}
\end{minipage}

\vspace{2mm}
\noindent
\begin{minipage}[H]{0.72\textwidth}
{\bf Collision of associate messages}. Once a tentative cluster head reaches the timeout without finding a closer cluster head, it returns to the parent channel and sends an associate message to the parent. However, in the physical topology given here it is likely that two nodes send associate messages at the same time to the same cluster head. This leads to a collision, and in such a case the specification does not state how the tentative cluster heads should proceed.
\end{minipage}
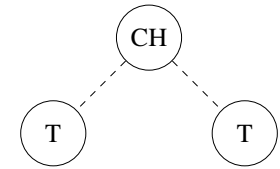
\begin{minipage}[H]{0.3\textwidth}
\centering
\scalebox{0.9}{
\begin{tikzpicture}[shorten >=1pt,node distance=2cm,on grid,auto]
   \node[state] (q_0)   {CH};
   \node[state] (q_1) [below right=of q_0] {T};
   \node[state] (q_2) [below left=of q_0] {T};
    \path[dashed,line width=0.4pt]
    (q_0) edge node {} (q_1)
          edge node {} (q_2);
\end{tikzpicture}}
\end{minipage}

\vspace{2mm}
\noindent
\begin{minipage}[H]{0.72\textwidth}
{\bf Narrow bridge problem}. Consider the physical topology to the right, in which the connection of the original cluster head to the rest of the nodes in the network is only possible through a close node. In this case, the direct child of the cluster head becomes a cluster slave. Subsequently since the cluster slave does not send out \texttt{BEACON} messages in its vicinity, the rest of the network never joins the tree. This problem arises due to the fact that the condition for becoming a cluster slave does not take the importance of the position of a node into account.
\end{minipage}
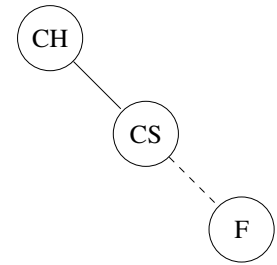
\begin{minipage}[H]{0.3\textwidth}
\centering
\scalebox{0.9}{
\begin{tikzpicture}[shorten >=1pt,node distance=2cm,on grid,auto]
   \node[state] (q_0)   {CH};
   \node[state] (q_1) [below right=of q_0] {CS};
   \node[state] (q_2) [below right=of q_1] {F};
    \path[-,line width=0.4pt]
    (q_0) edge node {} (q_1);
    \path[dashed,line width=0.4pt]
    (q_1) edge node {} (q_2);
\end{tikzpicture}
}
\end{minipage}

\vspace{2mm}

For certain topologies, the narrow bridge problem is inevitable for every possible execution. Moreover, given some fixed topologies and the initial conditions in which all nodes start on the same channel, collision of acknowledgements is also inevitable for every possible execution. This means that in these cases, there is no execution of the protocol in which a full cluster-tree is achieved. 

\subsection{Scalability of the cluster-tree formation protocol}\label{sec:scalability}
The main problem with the cluster-tree formation protocol however is its performance. In this section we give an argument for how the performance scales when the number of nodes in the network grows, by finding lower bounds on the termination time of the cluster-tree formation protocol. In order to proceed with our investigation we consider topologies in which the narrow bridge problem does not arise, and we solve the issue of collision of acknowledgements by changing the protocol as follows. After a free node receives a \texttt{BEACON} message, it does not acknowledge it, but instead it generates a random waiting time. If the cluster head is close, the free node waits for the random time and then it associates with the cluster head and becomes a cluster slave. Otherwise, it directly changes its type to tentative and follows the procedure we earlier specified.

\begin{figure}
\centering
\begin{subfigure}{0.30\textwidth}
\scalebox{0.8}{
\begin{tikzpicture}[level/.style={sibling distance=20mm/#1}]
\node at (0,1) {};
   \node[circle,draw] (a){1}
   	child {node[circle,draw] (aa) {2}}
	child {node[circle,draw] (ab) {3}}
   ;
\node at (0,-4.05) {};
\node at (-2.3,-0.95) {};
\end{tikzpicture}
}
\caption{A simple network of 3 nodes, with node 1 the original cluster head and node 2 and 3 free nodes.}\label{fig:abintree}
\end{subfigure}
\begin{subfigure}{0.20\textwidth}
\end{subfigure}
\begin{subfigure}{0.60\textwidth}
\scalebox{0.75}{
\begin{tikzpicture}[level/.style={sibling distance=60mm/#1}]
   \node[circle,draw] (a){1}
   	child {node[circle,draw] (aa) {2}
   		child {node[circle,draw] (aaa) {3}
			child[fill=none] {edge from parent[draw=none]} 
   		}
   		child {node[circle,draw] (aab) {5}
			child {node[circle,draw] (aaba) {6}}
   			child {node[circle,draw] (aabb) {7}}   
   		}
   	}
	child {node[circle,draw] (ab) {8}
		child {node[circle,draw] (aba) {9}
   			child {node[circle,draw] (abaa) {10}}
   			child[fill=none] {edge from parent[draw=none]} 
   		}
   		child {node[circle,draw] (aab) {11}
			child {node[circle,draw] (abba) {12}}
			child {node[circle,draw] (abbb) {13}}
   		}	
	}
   ;
   \draw [dashed] plot [smooth cycle] coordinates {(0,0.5) (-3.6,-1.8) (3.6,-1.8) };
	\draw [dashed] plot [smooth cycle] coordinates {(-3,-1.0) (-5,-3.3) (-1,-3.3) };
	\draw [dashed] plot [smooth cycle] coordinates {(3,-1.0) (5.1,-3.3) (1,-3.3) };
	\draw [dashed] plot [smooth cycle] coordinates {(-1.5,-2.5) (-3.1,-4.8) (0,-4.8) };
	\draw [dashed] plot [smooth cycle] coordinates {(1.5,-2.5) (0,-4.8) (3,-4.8) };
	\draw [dashed] plot [smooth cycle] coordinates {(4.5,-2.5) (3,-4.8) (6,-4.8) };
\end{tikzpicture}
}
\caption{A network with a balanced binary tree topology with height $h=3$. The number of nodes on level $i$ of the tree is $2^i$, except for $i=3$. The triples used in our scalability argument are encircled by dashed lines.}\label{fig:bbintree}
\end{subfigure}
\caption{Scalability analysis of a network with 3 nodes and a network corresponding to a binary tree.}
\end{figure}
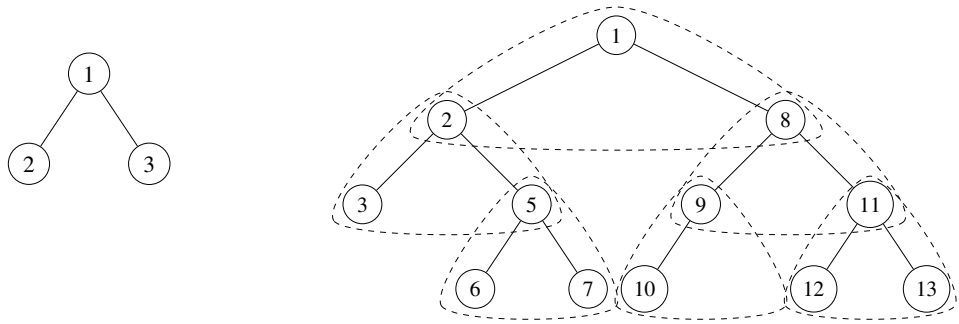

In the next step we take some of the witnesses of property (\ref{eq:success}), i.e., paths in the state space over which the property holds, and we calculate the amount of time it takes for the procedure to complete in a network of three nodes, with the physical topology given in Figure~\ref{fig:abintree}. In order to find witnesses we use the lpsxsim tool, which given a linearised mCRL2 process specification allows the manual generation of traces. Using this tool and our knowledge of the protocol we look for the smallest sequence of actions for which property (\ref{eq:success}) is satisfied. 

Let $\texttt{MAX\_ID}=3$, $\texttt{NUM\_OF\_CHANNELS}=3$ and the minimum time for nodes to remain tentative be 2. We find an initial configuration for which the network model satisfies property (\ref{eq:success}), and has the smallest trace for a witness. In the witness we find the minimum time for the formation of the cluster is 8 time-slots. In the setting of a slotframe which reserves 2 out of 12 time-slots for cluster-tree formation and maintenance, this means that the cluster-tree formation takes at least 3 slotframes to complete. Suppose that a time-slot is 120 milliseconds long, then it takes 5760 milliseconds for the algorithm to form a tree.

We now give our argument on how performance scales for larger networks. First, consider a tentative node with $\textit{id}=A$ which wants to join a cluster head in tier $t$. Since every association needs to be handled by the original cluster head, the cluster head on tier $t$ has to forward the association request to its parent on tier $t-1$, and so forth until it reaches the original cluster head. In the same manner the association response needs to travel back to node $A$. However, in case other requests arrive at a cluster head which is involved in this process at the same time, the performance of the protocol quickly deteriorates. Therefore for a node at tier $t$ the association procedure takes at least $2t$ to complete.

For simplicity consider the physical topology to be according to a balanced binary tree of height $h$. An example of such a network is given in Figure~\ref{fig:bbintree}, in which node $1$ is the original cluster head. The edges in this tree indicate the pairs of nodes that are in range of each other. The tree structure resulting from the protocol in such a case is identical to the physical topology.

We decompose the balanced binary tree into triples of nodes, as shown in Figure~\ref{fig:bbintree}. Let $0<i<h$, each triple consists of a parent with node height $i-1$ and its children with node height $i$. Consider a point in the execution of the cluster-tree formation protocol in which the parent node has already joined the network but its two children are still free. Using the witness we derived earlier, we see that excluding the messages \texttt{ASSOCIATE} and \texttt{ASSOCIATE\_ACK} the minimum time required for message exchange inside the triple (\texttt{BEACON} message and tentative waiting time) is 5 time-slots. Therefore, following our earlier observation for a single node with node height $i$, the time for the completion of the procedure is $5+2i$, and for two nodes this time is $5+4i$. 

Consider the ideal case in which: all messages avoid collisions, every cluster head is involved in at most one association at a time and every free node listens on the channel of its future parent exactly as soon as the parent sends a beacon. Since in a balanced binary tree the number of nodes with node height $i<h$ is exactly $2^i$ the association of all the nodes with height $i$ takes $5+2^{i+1}i$ time-slots to complete. 

There is one exception for simultaneous handling of associations which happen in the two separate subtrees of the original cluster head. In this case two associations which arrive at the original cluster head only within $2$ time-slots from each other can be processed by the original cluster head. By using this insight the minimum time for association of all nodes of height $i$ becomes $7+2^ii$.

And lastly, we know that a balanced binary tree of height $h$ has at least 1 node with node height $h$, therefore the lower bound on the number of time-slots needed for the formation of the cluster-tree in such a topology is given by the formula:
\begin{equation}\label{eq:time1}
5+2h+\sum_{i=1}^{h-1}\left(7+2^{i}i\right)
\end{equation}

As an example, consider a network with 500 nodes. A balanced binary tree with 500 nodes has height $h=8$. This means that the formation of the cluster-tree in such a network takes at least $27$ minutes to complete in the extremely ideal case. With the introduction of collisions, random back-offs, and the increase of the number of channels this time increases substantially.

Based on the lower bound given by formula~(\ref{eq:time1}), we conclude that for a network of $n$ nodes the optimal time for the execution of the cluster-tree formation protocol is $\Omega(n\log(n))$, and hence super linear.

\section{Further details on modelling and verification}
The activities of modelling and verification of the cluster-tree formation protocol was done in a period of five weeks, with approximately 
100 hours of effort. This included becoming familiar with the domain and consultation with the designers and domain experts. 
The generated state spaces are so small that tool effort for verification is negligible.

Our models and the modal formula are included in the model repository:
\begin{itemize}
\setlength\itemsep{-0.2ex}
\item[-] \texttt{ClusterFormation.mcrl2}: the mCRL2 specification corresponding to Sections~\ref{sec:informal} and~\ref{sec:twolayer}.
\item[-] \texttt{ClusterFormation\_NoAcks.mcrl2}: the mCRL2 specification of the protocol with no \texttt{BEACON} acknowledgements, as described in Section~\ref{sec:scalability}.
\item[-] \texttt{property1.mcf}: the correctness property of the protocol (formula (\ref{eq:success})).
\end{itemize}


\section{Conclusion}\label{sec:conc}
In this paper we showed how specification and verification of a network protocol by the mCRL2 language and toolset can give insights
to the operation of wireless network protocols. Remarkably, by only modelling a few number of nodes (3 nodes in our case) we 
were able to detect issues in the protocol design such as scalability. The key message here is that by modelling on a suitable 
abstraction level and with a sufficient number of components, one can effectively reason about the performance and correctness 
of wireless network protocols.

Based on our findings, we formulate a number of guidelines which can help in the general design of protocols using the 
IEEE 802.15.4 TSCH MAC layer operation mode:
\begin{itemize}
\setlength\itemsep{-0.2ex}
\item Broadcast messages should not be acknowledged (which is already common knowledge).
\item Any other sort of communication triggered by single shared events (like a broadcast), should use back-off schemes or channel sensing strategies to promote robustness.
\end{itemize}
The following guidelines are specific to the design of an improved cluster-tree formation protocol:
\begin{itemize}
\setlength\itemsep{-0.2ex}
\item A node should collect information regarding its neighbours before becoming a cluster slave.
\item To achieve better performance, the cluster-tree formation should be done during a dedicated, intensive period with no slotframe, in which only cluster-tree formation messages are exchanged.
\item The association requests should not be forwarded one by one to the original cluster head.
\end{itemize}

\nocite{*}
\bibliographystyle{eptcs}
\bibliography{all}

\begin{thebibliography}{10}
\providecommand{\bibitemdeclare}[2]{}
\providecommand{\surnamestart}{}
\providecommand{\surnameend}{}
\providecommand{\urlprefix}{Available at }
\providecommand{\url}[1]{\texttt{#1}}
\providecommand{\href}[2]{\texttt{#2}}
\providecommand{\urlalt}[2]{\href{#1}{#2}}
\providecommand{\doi}[1]{doi:\urlalt{http://dx.doi.org/#1}{#1}}
\providecommand{\bibinfo}[2]{#2}

\bibitemdeclare{misc}{mCRL2}
\bibitem{mCRL2}
\emph{\bibinfo{title}{{mCRL2} Homepage}}.
\newblock \bibinfo{howpublished}{\url{http://www.mcrl2.org/}}.
\newblock \bibinfo{note}{Accessed: 2017-01-13}.

\bibitemdeclare{article}{TSCH2012}
\bibitem{TSCH2012}
 (\bibinfo{year}{2012}): \emph{\bibinfo{title}{{IEEE Standard for Local and
  Metropolitan Area Networks\textemdash Part 15.4: Low-Rate Wireless Personal
  Area Networks (LR-WPANs)}, Amendment 1: MAC Layer}}.
\newblock {\sl \bibinfo{journal}{IEEE Computer Society}}
  \bibinfo{volume}{2012}, \doi{10.1109/IEEESTD.2012.6185525}.

\bibitemdeclare{article}{TSCH2015}
\bibitem{TSCH2015}
 (\bibinfo{year}{2016}): \emph{\bibinfo{title}{{IEEE 802.15.4-2015 - IEEE
  Standard for Low-Rate Wireless Networks}}}.
\newblock {\sl \bibinfo{journal}{IEEE Computer Society}},
  \doi{10.1109/IEEESTD.2016.7460875}.

\bibitemdeclare{inproceedings}{dandelski2016rll}
\bibitem{dandelski2016rll}
\bibinfo{author}{Conrad \surnamestart Dandelski\surnameend},
  \bibinfo{author}{Bernd-Ludwig \surnamestart Wenning\surnameend},
  \bibinfo{author}{Michael \surnamestart Kuhn\surnameend} \&
  \bibinfo{author}{Dirk \surnamestart Pesch\surnameend} (\bibinfo{year}{2016}):
  \emph{\bibinfo{title}{{RLL-reliable low latency broadcast data dissemination
  in dense wireless lighting control networks}}}.
\newblock In: {\sl \bibinfo{booktitle}{Emerging Technologies and Factory
  Automation (ETFA), 2016 IEEE 21st International Conference on}},
  \bibinfo{organization}{IEEE}, pp. \bibinfo{pages}{1--8},
  \doi{10.1109/ETFA.2016.7733751}.

\bibitemdeclare{article}{802.15.4e-survey}
\bibitem{802.15.4e-survey}
\bibinfo{author}{Domenico \surnamestart De~Guglielmo\surnameend},
  \bibinfo{author}{Simone \surnamestart Brienza\surnameend} \&
  \bibinfo{author}{Giuseppe \surnamestart Anastasi\surnameend}
  (\bibinfo{year}{2016}): \emph{\bibinfo{title}{{IEEE 802.15. 4e: A survey}}}.
\newblock {\sl \bibinfo{journal}{Computer Communications}}
  \bibinfo{volume}{88}, pp. \bibinfo{pages}{1--24},
  \doi{10.1016/j.comcom.2016.05.004}.

\bibitemdeclare{book}{fokkink2013distributed}
\bibitem{fokkink2013distributed}
\bibinfo{author}{Wan \surnamestart Fokkink\surnameend} (\bibinfo{year}{2013}):
  \emph{\bibinfo{title}{Distributed Algorithms: An Intuitive Approach}}.
\newblock \bibinfo{publisher}{The MIT Press}.

\bibitemdeclare{article}{gallager1983distributed}
\bibitem{gallager1983distributed}
\bibinfo{author}{Robert~G. \surnamestart Gallager\surnameend},
  \bibinfo{author}{Pierre~A. \surnamestart Humblet\surnameend} \&
  \bibinfo{author}{Philip~M. \surnamestart Spira\surnameend}
  (\bibinfo{year}{1983}): \emph{\bibinfo{title}{A distributed algorithm for
  minimum-weight spanning trees}}.
\newblock {\sl \bibinfo{journal}{ACM Transactions on Programming Languages and
  systems (TOPLAS)}} \bibinfo{volume}{5}(\bibinfo{number}{1}), pp.
  \bibinfo{pages}{66--77}, \doi{10.1145/357195.357200}.

\bibitemdeclare{article}{garay1998sublinear}
\bibitem{garay1998sublinear}
\bibinfo{author}{Juan~A \surnamestart Garay\surnameend}, \bibinfo{author}{Shay
  \surnamestart Kutten\surnameend} \& \bibinfo{author}{David \surnamestart
  Peleg\surnameend} (\bibinfo{year}{1998}): \emph{\bibinfo{title}{A sublinear
  time distributed algorithm for minimum-weight spanning trees}}.
\newblock {\sl \bibinfo{journal}{SIAM Journal on Computing}}
  \bibinfo{volume}{27}(\bibinfo{number}{1}), pp. \bibinfo{pages}{302--316},
  \doi{10.1137/S0097539794261118}.

\bibitemdeclare{techreport}{grieco2015using}
\bibitem{grieco2015using}
\bibinfo{author}{L~\surnamestart Grieco\surnameend} (\bibinfo{year}{2015}):
  \emph{\bibinfo{title}{{Using IEEE 802.15. 4e Time-Slotted Channel Hopping
  (TSCH) in the Internet of Things (IoT): Problem Statement}}}.
\newblock \bibinfo{type}{Technical Report}, \doi{10.1109/IEEESTD.2012.6185525}.

\bibitemdeclare{book}{groote2014modeling}
\bibitem{groote2014modeling}
\bibinfo{author}{Jan~Friso \surnamestart Groote\surnameend} \&
  \bibinfo{author}{Mohammad~Reza \surnamestart Mousavi\surnameend}
  (\bibinfo{year}{2014}): \emph{\bibinfo{title}{Modeling and analysis of
  communicating systems}}.
\newblock \bibinfo{publisher}{The MIT press}.

\bibitemdeclare{inproceedings}{gross2007does}
\bibitem{gross2007does}
\bibinfo{author}{Christian \surnamestart Gro{\ss}\surnameend},
  \bibinfo{author}{Holger \surnamestart Hermanns\surnameend} \&
  \bibinfo{author}{Reza \surnamestart Pulungan\surnameend}
  (\bibinfo{year}{2007}): \emph{\bibinfo{title}{Does clock precision influence
  ZigBee’s energy consumptions?}}
\newblock In: {\sl \bibinfo{booktitle}{International Conference on Principles
  of Distributed Systems}}, \bibinfo{organization}{Springer}, pp.
  \bibinfo{pages}{174--188}, \doi{10.1007/978-3-540-77096-1\_13}.

\bibitemdeclare{inproceedings}{kauer2016formal}
\bibitem{kauer2016formal}
\bibinfo{author}{Florian \surnamestart Kauer\surnameend},
  \bibinfo{author}{Maximilian \surnamestart K{\"o}stler\surnameend},
  \bibinfo{author}{Tobias \surnamestart L{\"u}bkert\surnameend} \&
  \bibinfo{author}{Volker \surnamestart Turau\surnameend}
  (\bibinfo{year}{2016}): \emph{\bibinfo{title}{{Formal Analysis and
  Verification of the IEEE 802.15. 4 DSME Slot Allocation}}}.
\newblock In: {\sl \bibinfo{booktitle}{Proceedings of the 19th ACM
  International Conference on Modeling, Analysis and Simulation of Wireless and
  Mobile Systems}}, \bibinfo{organization}{ACM}, pp. \bibinfo{pages}{140--147},
  \doi{10.1145/2988287.2989148}.

\bibitemdeclare{book}{lynch1996distributed}
\bibitem{lynch1996distributed}
\bibinfo{author}{Nancy~A \surnamestart Lynch\surnameend}
  (\bibinfo{year}{1996}): \emph{\bibinfo{title}{Distributed algorithms}}.
\newblock \bibinfo{publisher}{Morgan Kaufmann}.

\end{thebibliography}

\end{document}